\documentclass[letterpaper]{JHEP3}

\usepackage{latexsym}
\usepackage{amsmath, mathrsfs}
\usepackage{graphicx, slashed}

\title{Toptet}

\author{Thomas Gr\'egoire\\
        Ottawa-Carleton Institue for Physics, Physics Department, Carleton University, Ottawa, ON, Canada. \email{gregoire@physics.carleton.ca}}
\author{Emanuel Katz \\
	Physics Department,  Boston University, Boston, MA, USA. 
	\email{amikatz@buphy.bu.edu}}        
\author{Veronica Sanz\\
        Department of Physics and Astronomy, York University, Toronto, ON, Canada. \email{vsanz@yorku.ca} }

\abstract{
Final states with four tops appear in various extensions of the Standard Model.  Alas, top reconstruction faces combinatorial issues as they show up as large multiplicity events. In this paper, we present a new procedure to determine whether new physics 
is in fact due to a new source for tops.  We establish the use of this procedure to separate the signal from background
(primarily $t\bar{t}$+jets).  Our analysis is model independent, in that it does not use any details of the four top production (such as
possible missing energy), and does not require b-tagging.  
}

\preprint{}

\newcommand{\be}{\begin{equation}}
\newcommand{\ee}{\end{equation}}
\newcommand{\bea}{\begin{eqnarray}}
\newcommand{\eea}{\end{eqnarray}}
\newcommand{\gev}{\textrm{ GeV}}

\newcommand{\ttb}{$t \bar{t}$ }
\newcommand{\MET}{E_T\hspace{-0.230in}\not\hspace{0.18in}}

\begin{document}
\section{Introduction}

One of the central questions to be addressed by the Large Hadron Collider is whether electroweak symmetry breaking (EWSB) 
occurs within a natural solution to the hierarchy problem.
In many natural theories of electroweak breaking the top plays a special role.  Indeed, its large
mass, and therefore strong coupling to the Higgs suggests that it may have significant interactions
with the new physics which leads to electroweak symmetry breaking (EWSB).  Thus, if we should see new
particles at the LHC, it may be likely that these particles decay into final states containing the
top at a significant rate.  A possible litmus test for a naturalness is therefore the presence of new
sources for tops.  The fact that the top carries color makes it even more likely
that it will appear in the decays of new colored particles.  Since the LHC is effectively a gluon collider,
the new particles it will produce will predominantly be colored.  New physics at the LHC thus promises
to be rich source for tops and possibly bottoms.  On the other hand, standard model direct top production will also be
significant at the LHC, which is sometimes referred to a \emph{a top factory}.  The 
standard model (SM) top background is therefore a formidable challenge for discovering new top sources.
In some cases, this challenge can be met as, for example, in models where tops originating from decays are
very boosted.  In such cases, one can reduce the background by studying ``top-jets'' \cite{boosted}.  Here we will instead 
focus on scenarios where one readily gets four tops in the final state\cite{fourtops}.  
Such scenarios arise naturally both in composite Higgs models, as well as in certain SUSY models.
These events are interesting in that they contain a large multiplicity
of particles.  Indeed, one typically gets many jets, including several b-jets and leptons.
Thus, discovering the new physics should not be difficult.  The challenge lies in identifying the
origin of the signal as coming from top quarks while at the same time beating SM background.  
Establishing the link of the new physics to tops is, however, very important as that will provide direct clues  
about electroweak breaking.  In this paper we will show that it is possible to determine that
the new physics is associated with tops.  The procedure is to first to reconstruct tops hadronically,
and to count the number of events which contain multiple top candidates and a lepton.  With a sufficiently 
restrictive cut, it is possible to beat the primary background, which consists of $t\bar{t} + jets$.
This is possible for $\sim 10 ~pb$ four top production cross sections at $\sim 1 fb^{-1}$.
The "topness" measurement consists of first of seeing a significant number of events which contain
two hadronic tops, as well as seeing the variation of the number of such events,
as a function of the mass parameter, $m_t$, used in the top reconstruction.  
We show that number of events containing two tops peaks near the physical value,
$m_t=175$, for two different top reconstruction variables.

The paper is organized as follows.  In section \ref{toprich} we will summarize possible scenarios where four top 
production is possible at a significant rate.  Section \ref{counting} is a description of the counting procedure and 
top selection criteria.  In section \ref{varying} we apply our selection criteria to count tops in four top final states, and 
discuss when four tops rise above SM background.  We then describe the effect of varying the top reconstruction
mass parameter.  We conclude in section \ref{conclusions}.

\section{Top rich scenarios} \label{toprich}
The third generation of quarks play a special role in many scenarios of new physics, specially if those proposals address questions such as Higgs tuning or compositeness. In  this section, we sketch some scenarios where tops are produced copiously, from supersymmetry with light stops to models of strong electroweak symmetry breaking.  Our focus here is only to mention possible 
top rich scenarios, however we will not be attempting to distinguish them from each other. 

\subsection{Composite Higgs}
In beyond the SM scenarios which address the hierarchy problem, the top will have partners that
remove the quadratically divergent contribution to the Higgs mass from top loops.  In models where
the Higgs is composite \cite{compositeness}, the top must also be largely composite to account for the top Yukawa.
The top partners, in this case, are new composite colored particles.  The top obtains its large mass 
through mixing with these composite particles, which in turn couple strongly to the Higgs.  The top 
(and possibly the left-handed bottom) thus develop significant interactions with the composite sector.
Hence, final states containing tops and bottoms offer a special window into new physics.  

At the LHC, the most promising signals will come from production of new colored particles, which will be made with a
significant cross section.  The types of new colored composite particles to be found at the LHC depend somewhat
on the details of the model.  However a minimal choice is that the left-handed quark doublet, $q$, 
and the right-handed top, $\bar{t}$, mix with three fermion composite states of the form
\begin{equation}
q \sim (\psi_3 ~\psi_2 ~\chi), ~\bar{t} \sim (\psi_{\bar 3} ~\psi_0 ~\chi),
\end{equation}
where $\psi_3$ is a color triplet, $\psi_2$ is a weak doublet, and $\chi$ a SM singlet.  
Rearranging the preon fermions therefore gives the additional composite fermions:

\begin{equation}
\tilde{g} \sim (\psi_3 ~\psi_{\bar 3} ~\chi), ~Q \sim (\psi_3 ~\psi_3 ~\chi),~\bar{Q} \sim (\psi_{\bar 3} ~\psi_{\bar 3} ~\chi).
\end{equation}

The first has the quantum numbers of the gluino, while the second set (which form a Dirac fermion) can be either a 
fundamental or a six of color, depending on the confining gauge group of the composite sector.
In the following we will assume the $\bar{Q}$ to be a color triplet for simplicity, but we do not expect the resulting 
phenomenology to depend significantly on the this choice.  Similarly, one can also have composite fermions with 
the quantum numbers of Binos, Higgsinos, and Winos (again depending on the gauge group of the composite sector).  
On general grounds, the above fermions will have interactions of the type
\begin{equation}
\it{L} = c_1 \frac{\tilde{g} t\bar{t} N}{f^2} + c_2 \frac{Q \bar{t}\bar{t} N}{f^2}
 + c_3 \frac{\tilde{g} b \bar{t} C}{f^2} + c_4 \frac{Q \bar{t} \bar{b}\bar{C}}{f^2} + h.c, 
\end{equation}
where $C$ and $N$ are the lightest charged and neutral composite fermions, respectively. In writing these interactions we take
the new particles to a parity symmetry (similar to R-parity of SUSY), which occurs naturally in composite models. 
The scale $f$ is typically of order 1 TeV, and the $c_i$ generally depend on various mixing angles.  There is a significant region of 
parameter space in these models where the mixing between the quark doublet, $q$, and the composite particles is smaller
than that of the right-handed top singlet, $t$.  Thus, the last two terms will be smaller than the first two, and as long
as the spectrum allows it, a decay of the sort
\begin{equation} 
\tilde{g} \rightarrow t\bar{t}N
\end{equation}
occurs at a significant rate.  Since the gluinos will be pair produced,  one has events with four tops.   Pair production of
$Q$'s will similarly lead to a four top final state.  Since $\tilde{g}$ is a color octet, it will generically have a larger
production cross section, and we therefore assume for simplicity that it constitutes the predominant new source for tops.

\subsection{Supersymmetric models}

One can also have SUSY scenarios which are top rich.  As bounds on flavor changing neutral currents (FCNC) are most stringent on the first two generations, there are a variety of models where the superpartners of the first two generations are heavy.
The left-handed and right-handed stops, on the other hand, cannot be too heavy without significantly increasing the fine tuning of the Higgs mass.  The Higgs mass is also sensitive to contribution from the masses of the weak gauginos and the higgsinos.
Thus, some natural models without FCNC include spectra where the light sparticles are gauginos, higgsinos, and third generation 
squarks \cite{stops}.  A cascade of the following form can then occur at a significant rate (especially if the lightest squark is the right-handed stop)

\begin{equation} 
\tilde{g} \rightarrow \tilde{t} \bar{t} \rightarrow t \bar{t} \tilde{B}
\end{equation} 

Again, as with the composite models, one then gets events with four tops in the final state.  

\subsection{Vectorlike Confinement} 
 
 Another interesting context for multiple top production are the recently proposed vectorlike models~\cite{vectorlike}, where one has a new strongly coupled sector at a TeV,  which however is not involved in electroweak breaking.  Here, one has new vector-boson state, the hyper-$\rho$, and new pseudoscalar states, the hyper-pions, all bound states of hyper-quarks which are charged under the standard model and also under the new strong interactions.  The phenomenology of the strongly coupled sector is modeled after QCD.  Thus,
 the hyper-$\rho$ may decay into two hyper-pions.  The hyper-pions
 which are charged under the SM gauge groups, will decay to SM fermions via 4-Fermi operators coupling hyper-quarks, $\Psi$, 
 to regular quarks.  One can imagine that due to Flavor Changing Neutral Current constraints, the 4-Fermi operators
 can be the largest for the third generation,
 \begin{equation}
 \it{L} = \frac{\bar{\Psi}_{\bar{i}} \gamma_5\gamma_\mu\Psi_j \bar{t}_R^{\bar{j}} \sigma^\mu t_R^i}{M^2}.
 \end{equation}
 Here $i,j$ are color indices. The  colored hyper-pions, $\tilde{\pi}_{\bar{i}j}$, will then prefer to decay to tops.  
Direct colored hyper-pion production will then be a source of four tops.
\section{A measure of topness in the event} \label{counting}
A top decays into three objects, either 3 jets or $\ell+j+\MET$. In a multi-top environment, the final state is quite messy, and reconstruction requires some understanding of the kinematics of the final states. In this paper, we are  focusing in {\it early discovery} where the parent particles have a large production cross-section and are therefore rather light. Kinematically,  this means that the tops would be  produced with little phase space. Since the top parent itself does not have particularly large momentum, the top decay products are not particularly boosted~\footnote{Boosted tops are an active area of research and this kinematic feature has  been thoroughly exploited~\cite{boosted}}.  In this section we describe a measurements which show sensitivity to 
a new source of tops in the signal, which can easily overcome SM backgrounds for a four top signal with a cross section
around 10 pb.

The measurement is the number of events containing two tops decaying hadronically in a sample which contains at least
one lepton and more than eight jets
\footnote{Considering leptonically decaying tops as well does not significantly enhance our two 
top new physics signal and we will not consider this possibility further.}.  It is also possible to beat SM background
by considering more leptons and only reconstructing one hadronic top.  However, asking for two hadronic
tops is much more effective at eliminating non-top related jets, and
allows for greater sensitivity to an additional test for the presence of tops that we will describe in section
\ref{varying}.  It is thus a better choice if the goal is not simply the discovery of a new source of jets and leptons,
but the determination that these come from a new source for four tops.

Before discussing the top reconstruction, let us first focus on the effect of 
certain basic cuts on the various SM backgrounds.  These cuts are

 \bea
 \label{basic}
& & n_{\ell}   \geqslant 1 ,  p_{T,\ell} \geqslant 20 \gev \,  |\eta|_{\ell} < 2.5 \ , \, \Delta R_{\ell,o} > 0.4 \\ \nonumber
& & p_{T,j} \geqslant 30 \gev  \ , \MET    \geqslant 20 \gev \ , |\eta|_j < 3.5 \ , \, \Delta R_{jj} > 0.4, 
\eea 
The cuts on leptons and missing energy reduces the QCD background to acceptable levels \cite{tdrs}.  This leaves 
\ttb+jets, $W$+jets and $Z$+jets as the primary backgrounds.  These get reduced further by requiring more than nine
jets passing the above criteria.  Let us first describe the effect of the basic cuts on these backgrounds.

\subsection{The effect of basic cuts on the background}

As mentioned, the most relevant SM backgrounds to multi-top final states are \ttb+jets, $W$+jets and $Z$+jets. 
The samples are generated with ALPGEN~\cite{ALPGEN} using MLM matching~\cite{MLM} at 14 TeV.  
 The \ttb sample has no generation cuts and it contains 3M events. The cross sections for \ttb+ (0,1,2) jets are (440,778,730) pb. Leptonic $W$+jets is generated with parton level cuts $p_{T,o}>$ 15 GeV, $|\eta_{o}|<$ 4., $\Delta R_{o,o}>$ 0.4 and $\MET>$ 75 GeV, where $o$ is an object, either lepton or jet, $o=j, \ell$. After cuts on number of leptons and jets, only $W$+ 2 or more jets contributes to the analysis. We generated (500K,1.2M) events of $W+$(2,3) jets with (740,390) pb of cross section at parton level. Finally, we also generated $Z+$jets but after $\MET$ and number of jet cuts, this background is irrelevant.
  
 The last two steps of the simulated backgrounds are showering and detector effects. Hadronizaton/showering is simulated with PYTHIAv6.4 \cite{pythia}. More jets coming from radiation will be incorporated to the sample after PYTHIA. Pretty Good Simulator (PGS) \cite{pgs} is used to introduce some detector effects such as particle identification efficiencies and energy-momentum smearing.

After the generation in PGS is done, we perform the high-level analysis with the cuts of \ref{basic}.


\subsection{Top reconstruction criteria and further background reduction}  
\label{ordering}

Our goal will be to look for events containing multiple hadronically decaying tops.
Our top candidates consist
of three jets, whose invariant mass is near the top mass, two of which have an invariant mass near the W mass.
We have also studied the effect of cutting on a different reconstruction variable (instead of the W mass requirement).  This 
variable, $c_{1b}$, is an angular variable defined in the top rest frame, and may be useful as a complimentary
check on the presence of tops.  We describe this variable below.

 

We do not use b-tagging as this keeps more signal events, and is not necessary to remove background.
However, when constructing top candidates we start with the highest $p_T$ jet which has a higher likelihood of
being a b-jet.  Our procedure is to cycle through all 3-jet combinations in a given event, always starting with
the highest momentum jet.  Should two top candidates share a jet, we pick the combination whose 
3-jet invariant mass is closest to the top mass, and discard the others. We also discard combinations with $m_{jb}> m_t$.
We refer to this as "ordering our top candidates". The procedure is then repeated until  we obtain a list of top candidates that do not share any jets, and are ranked by the closeness of their 3-jet invariant to the mass of the physical top.

\subsubsection{Background reduction through top ordering}
Below we present the effect of ordering on reducing the SM background.  After our top candidates have been ordered,
our top selection consists of taking either one or two of the leading top candidates for each event.
The number of events is normalized to a luminosity of 1 fb$^{-1}$ at energies of 14 TeV.
\bea
 \begin{array}{|c|c|c|c|}
 \hline \textrm{Sample} & \textrm{Total size} &  N_t \textrm{ asked for} &  \textrm{events left}   \\\hline t \, \bar{t} +  \textrm{jets} & 1.9 \times 10^6& 1 &  1.8  \times 10^5  \\\hline  t \, \bar{t} +  \textrm{jets}  & 1.9 \times 10^6 & 2  & 3 \times 10^3   \\\hline   W  +  \textrm{jets}  & 1.1 \times 10^6 & 1  & 7 \times 10^3     \\\hline W  +  \textrm{jets}  & 1.1 \times 10^6 & 2  & 37    \\\hline
  \end{array}
\eea

In the table below we show the relative efficiencies in the $N_t$=1 and 2 bins:
\bea
\begin{array}{|c|c|c|}
\hline \textrm{Sample} & N_t &  \epsilon_{\textrm{ordering}}  \\\hline  t \, \bar{t} +  \textrm{jets} & 1 & 9.4\% 
 \\\hline  W +  \textrm{jets} & 1 & 0.6\% 
 \\\hline   t \, \bar{t} +  \textrm{jets} & 2 & 0.15\% \\\hline  W +  \textrm{jets} & 2 & 0.003 \%  \\\hline  
\end{array}
\eea

These numbers indicate, roughly, that this method finds a fake top about $1\%$ of the time, and is able to 
find a true top about $10\%$ of the time.  We will see that requiring two tops is therefore already sufficient
to reduce the SM background to manageable levels, especially that we have not yet required any
significant $H_T$, missing $E_T$, or number of jet cuts. 




\section{Seeing tops in the four top signal}\label{varying}

In the previous section we saw that it is possible to suppress SM background by looking for events with two hadronic
tops in a sample of containing many jets and at least one lepton.  We now would like to study the effect of our cuts
on our four top signal.   In particular, since the main background is \ttb+jets,  as all SM background without tops are largely reduced by the selection procedure, we need to check that it can be made small compared to the signal.
 
Our analysis is model independent, and just assumes new physics has some branching ratio to $2 t + 2\bar{t}+X$. To simulate the  new physics signal, we generate a four top sample in a suspersymmetric cascade decay, namely pair production of gluinos which decay to stops and tops. The stops subsequently decay to more tops and the neutralino ($\tilde{g} \to t \tilde{t}$, and $\tilde{t} \to t \tilde{\chi}^0$).  This is a natural scenario if stops happen to be the lightest squarks. In this cascade, the final state consists on 
\bea
2 t + 2 \bar{t} +  \slashed{E}_T \, ,
\eea
As we want to keep the analysis as model independent as possible, we do not assume any missing energy distribution, or reconstruct supersymmetric masses. The sample is generated in Madgraph/MadEventv4.4.3 \cite{MG} at 14 TeV and the passed through PYTHIA and PGS in accordance with the background.  
The results shown here correspond to a 400 GeV gluino, with $\tilde{g} \tilde{g}$ production cross section of 12.6 pb at 14 TeV. Because our analysis requires very modest $p_T$ and missing energy cuts and no $H_T$ cut, the scaling with other energies and other masses is simple: from 7 to 14 TeV, the cross section decreases by a factor $\sim$ 13 and increasing the gluino mass from 400 GeV to 800 GeV decreases the cross section by a factor $\sim$ 60.

Note that our analysis is not tied to Supersymmetry (see Sec.~\ref{toprich} for an array of models with multi-top final states) and there are no studies on Supersymmetry with multi-top final states. But there are bounds on gluinos looking at multi-jet and b-jet final states in association with missing energy \cite{gluinos}. Those bounds do not apply to our scenario still we would like to mention that using these final states one could impose bounds on gluinos at about 700 GeV in the most optimal situation, but are considerably weakened in a squeezed spectrum. 



\subsection{New Physics versus SM tops}
 \begin{figure}[t]
\centering
\includegraphics[scale=0.3]{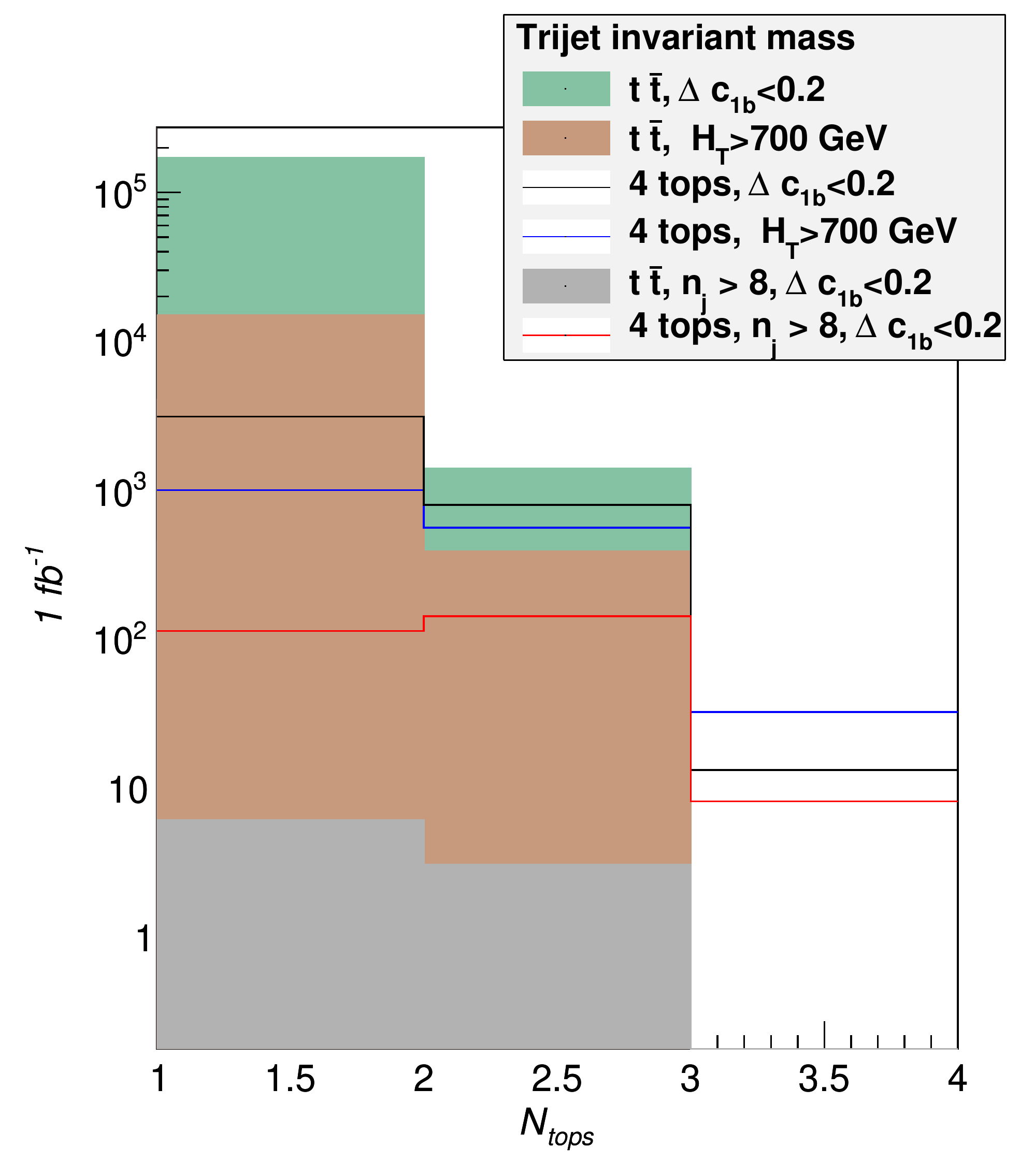}
\caption{The behavior of four-tops sample versus the \ttb+jets sample under different sets of cuts.}
\label{4tops}
\end{figure}   


In this section we present the effect of requiring two ordered tops on both the four top signal, and on the leading 
SM background ($t\bar{t}$+jets).  In addition to ordering the tops we will also require an additional cut
on the three jet invariant of $|m_{jjj} -m_t| < $ 30 GeV, which further helps properly group the jets.
We present numbers for 1 fb$^{-1}$ worth of data at 14 TeV, providing $1.3 \times 10^4$ four top events. 
Our results for number of events and efficiencies for different number of jets, $n_j$, are as follows:
\begin{center}
\bea
 \begin{array}{|c|c|c|c|c|c|}   \hline N_2=2 & 4 \textrm{ tops (eff.)} & t \bar{t}+\textrm{ jets (eff.)} & \frac{S}{B} & \frac{S}{\sqrt{B}} \\\hline
  \textrm{ordering}, \Delta m_{jjj}<30 ~GeV& 515 ~(4\%) & 1505 ~(.08\%) & 0.34 & 13 \\\hline   
  \textrm{and also} ~n_j > 6 & 487 ~(4\%) & 805 ~(.04\%) & 0.61  & 17 \\\hline   
  \textrm{and also} ~n_j > 8 & 229 ~(2\%) & 68 ~(.003\%) & 3.37 & 28   \\\hline \end{array} 
\eea
\end{center}
We note that these results do not require any significant missing $E_T$, and we have placed no cut on $H_T$ at all.
The last cut in the table appears to be especially efficient at reducing background.  As we have not placed any
$H_T$ cut, we can use the the efficiencies on the SM background, as an indication of the efficiencies
of our cuts on new physics as well.  It is clear that any new physics which is able to pass the last cut of the table must
have many jets and at least one lepton.  In addition, obtaining two fake tops from a sample which contains no tops
is a reduction by $\sim10^{-4}$, while finding two real tops is a reduction of $\sim .01$.   Thus, finding a percent efficiency
of a new physics signal to a two top requirement is evidence that the new physics indeed contains tops.  An additional litmus
test for "topness" will be discussed in the next section.

Let us also comment on the usefulness of a same-sign lepton (2SSL) cut, as in \cite{fourtops}.  This cut would lower \ttb substantially. We do not follow this procedure for two reasons: {\it 1.)} By asking for 2SSL, we reduce the number of tops we could reconstruct in the hadronic channels. The efficiency to reconstruct one top with a 2SSL cut is less than 5$\times 10^{-5}$, as compared with our proposed cuts' efficiency (at the few percent level), and {\it 2.)} Having only one reconstructed top, as opposed to two, makes the number of events less sensitive to varying the top mass reconstruction parameter (discussed below).  


\subsection{Varying $m_t$}

We now present an additional litmus test for the presence of tops.  The idea is to compare the content of the $N_t=2$ bin as the 
top mass parameter, $m_t$, is varied from $m_t=$155 GeV to 200 GeV.   In order to increase sensitivity to this parameter,
we require that our top candidates obey either $|m_{jj} -m_W| <$ 20 GeV (for two of the three jets), or that
a geometric parameter, $c_{jj}$ (described below), satisfy $|c_{jj} - c_{1b} (m_{jb})| < $ 0.1.


\subsubsection{Angular variables}

Let us now describe a set of variables that will prove useful for further purifying our top candidates.
Starting with a jet, which we will call a "b-jet" \footnote{Note, that we assume one of the jets to be a b-jet in our procedure, in hope that as we cycle trough our jets, the reconstruction will 
tend to identify b-jets, but we are not using actual b-tagging.}, 
we pair with 2 other jets in the event and find the following quantities
\bea
m_{jj}, \, m_{jb}, \, m_{jjb}, \, \cos(\theta_{jb}), p_{j}.
\eea
\begin{figure}[h!]
\centering
\includegraphics[scale=0.25]{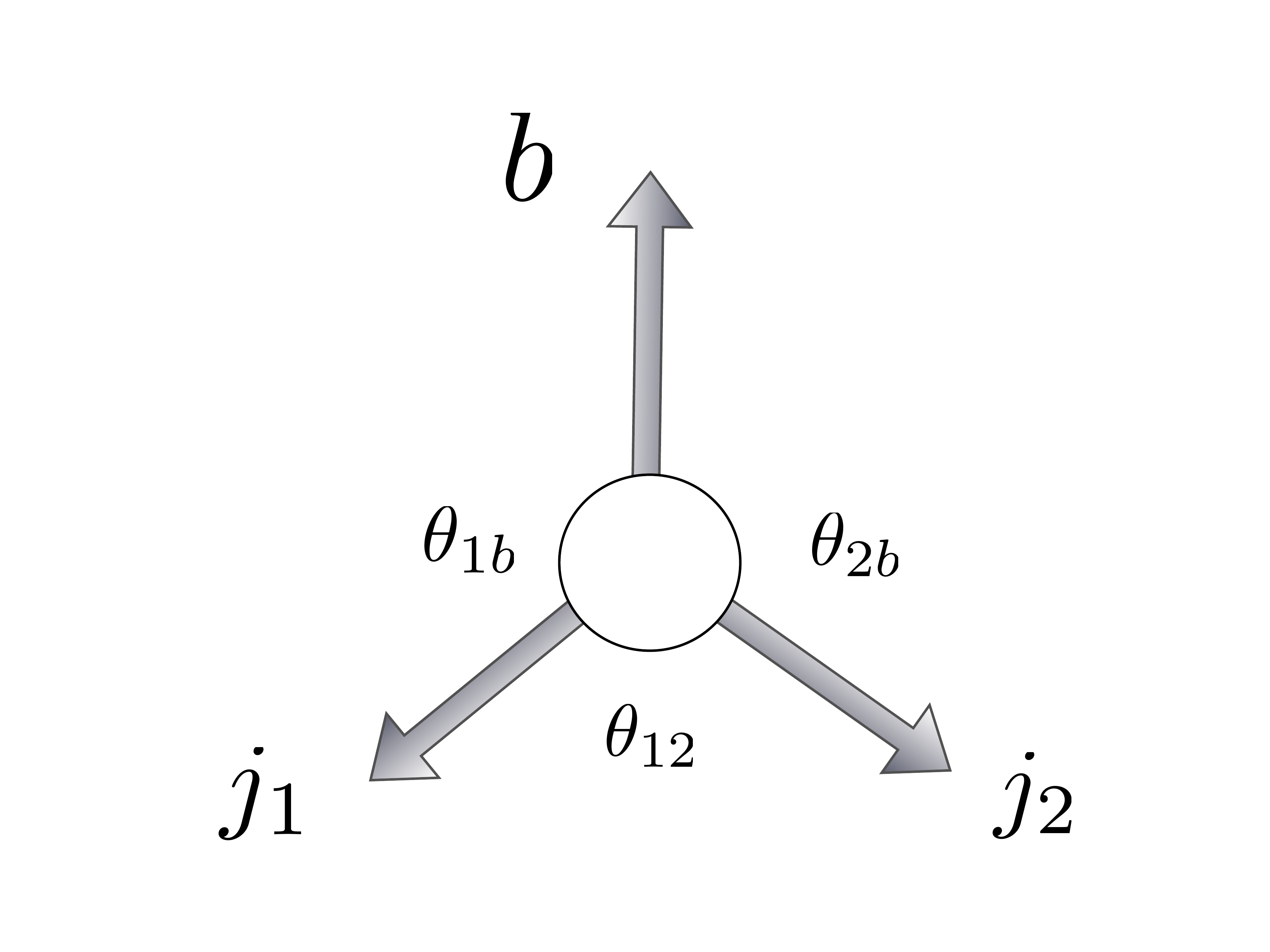}
\caption{The angular variables in the top rest frame.}
\label{topCM}
\end{figure}
The first three variables are standard.  The last set of variables consists of angles and momenta defined in the top rest frame as depicted in Fig.\ref{topCM}.  These are found by boosting the 3-jets to their center of mass frame.
For a proper identification of the jets coming from a particular top, these quantities must be kinematically related as:
\bea
p_1 &=& \frac{m_{1b}^2 + m_W^2}{2 m_t}\; \;, \\
p_2 &=&  \frac{m_t^2- m_{1b}^2}{2 m_t}\; \; ,\\
p_b &=&   \frac{m_t^2- m_W^2}{2 m_t}\; \; ,\\
\cos(\theta_{12}) &=& \frac{(m_{1b}^2 + m_W^2)(m_t^2- m_{1b}^2)-2m_W^2m_t^2}{(m_{1b}^2 + m_W^2)(m_t^2- m_{1b}^2)} \; \;,\\
\cos(\theta_{1b}) &=& \frac{(m_{1b}^2 + m_W^2)(m_t^2- m_W^2)-2m_{1b}^2m_t^2}{(m_{1b}^2 + m_W^2)(m_t^2- m_W^2)} \; \; .
\label{c1b}
\eea 
Thus, instead of demanding that that $m_{jj}$ be near $m_W$, we may require that for a given $m_{jb}$, one of the 
variables in \ref{c1b} be close to its kinematic value.

\subsubsection{Background reduction through top reconstruction}
We have studied the efficiency of applying cuts on the above rest frame variables and found that either $\cos(\theta_{1b})\equiv c_{1b}$ or $p_2$ were the best variables to discriminate between a signal with tops and backgrounds without tops.
In the following we will be using $c_{1b}$ to further purify our signal.   The numbers are normalized to $1.3 \times 10^4$ events.  Besides ordering, the previous cut on $\Delta m_{jjj}$, we have also applied a $n_{j}>$8 cut, and a cut on $c_{1b}$, which leads to 3 to 6 \ttb+jets background events at the maximum ($m_t=175$ GeV):

\begin{center}
\bea
 \begin{array}{|c|c|c|c|c}   \hline N_2=2 & 4 \textrm{ tops (eff.)} & t \bar{t}+\textrm{ jets (eff.)} & \frac{S}{B}\\\hline
 \textrm{previous cuts + } c_{1b}<0.1& 150 ~(1.15\%) & \sim 5 ~(\sim  0.0002\%) &  \sim 30\\\hline   
 \end{array} 
\eea
\end{center}

\begin{figure}[h!]
\centering
\includegraphics[scale=0.25]{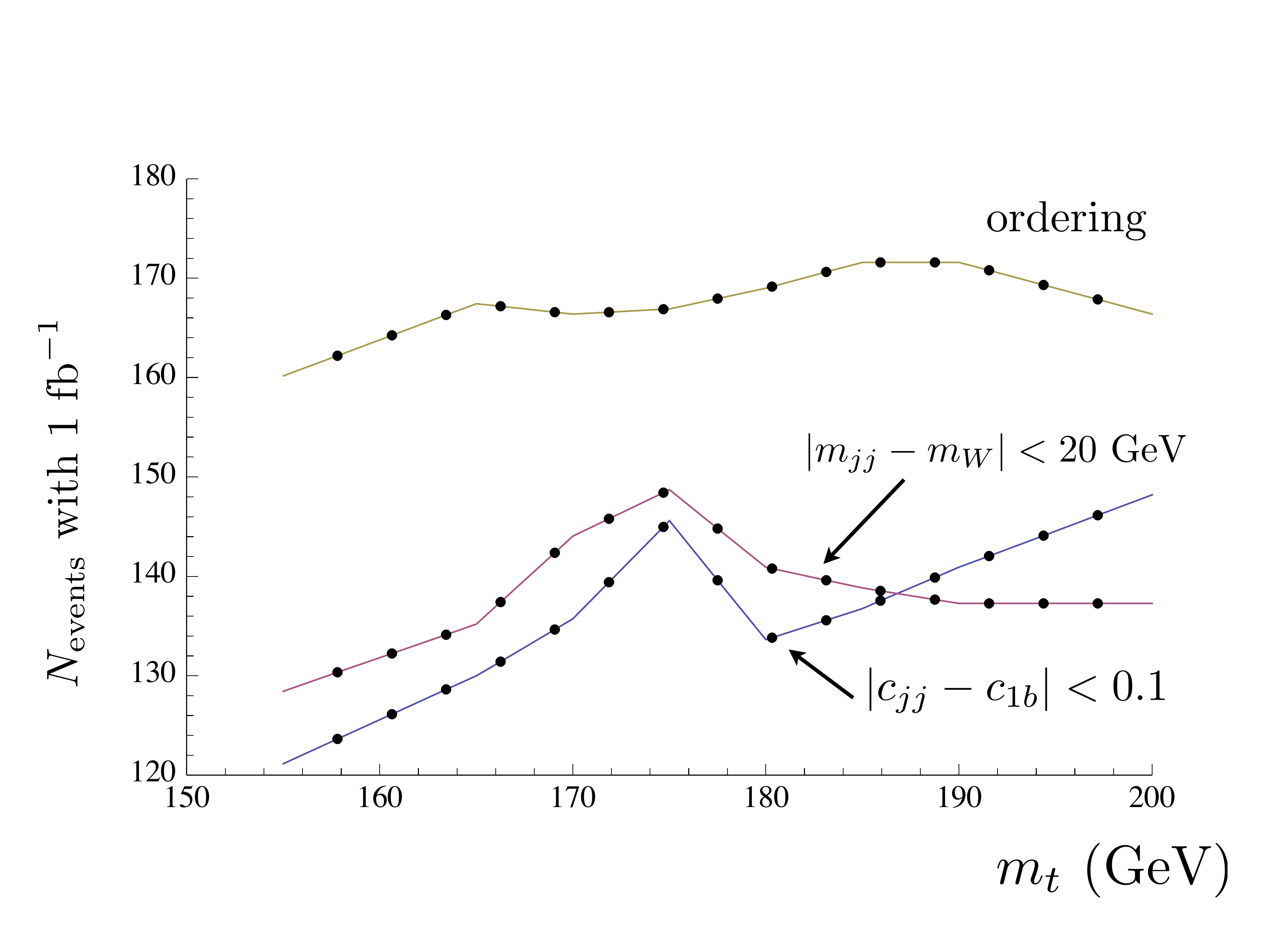}
\caption{The $N_t=2$ bin content as a function of $m_t$.}
\label{mtvary}
\end{figure}

This cut, although it reduces the signal further, has the advantage of nearly eliminating the background.  The sample
becomes sufficiently pure that we would expect sensitivity to the variation of the top mass parameter, $m_t$, of the
the reconstruction.  Indeed, we see this effect in Fig.  \ref{mtvary}, with the number of events peaking at the 
physical value of the top mass.  The figure also displays a similar effect when cutting on the W-mass.  However,
without either a $c_{1b}$ or a W-mass cut (i.e. with only ordering), there is little sensitivity  to the $m_t$ parameter.
\footnote{The poor behavior of $c_{1b}$ for larger values of $m_t$, can be understood from Eqn. (\ref{c1b}): 
$c_{1b}$ quickly asymptotes to  -1 for $m_t \gg m_W$, thereby effectively accepting any nearly back-to-back jets.}

One could, therefore, imagine looking for this sensitivity as an additional test for "topness". For example, a supersymmetric decay chain with charginos could produce some $W$'s ($\tilde{\chi}^{\pm} \to \tilde{\chi^0} W^{\pm}$). This kind of new physics would have a large efficiency to  the dijet invariant mass cut,  but would not show this dependence on $m_t$. 

To estimate the significance of the variation with $m_t$ one could vary the cuts and assign an error on $N_t$ based on that variation.  We think, however, that the sensitivity of $N_t$ with the value of $m_t$ can only be confirmed using a more realistic simulation and a larger sample of events.  For example, an important issue is the resolution on the $c_{1b}$ variable, which is obtained by boosting
from the lab frame to the top rest frame.   Thus, the energy-momentum resolution of the jets on the boost relate to the resolution on this variable.  We studied this by  matching parton-level events to PGS-level events in a 2-top sample, in a way very similar to \cite{topcms,tdrs}.  First we keep only events with $6,7$ or $8$ jets and match those jets to  partons in a way that minimizes the sum of the angles between a jet and its matching parton. We then require that all the partons in a top be within $15$ degrees of their matching jets. Finally, we require the parton-level tops to be within $15$ degrees of the matching PGS-level top.  
We fit the resulting spread in $c_{1b}$ to a gaussian, yielding the estimate 
\bea
c_{1b}-c_{1b}^{th} |_{\textrm{fit}}=  -0.007 \pm 0.128 \ .
\eea
This suggests that the effect of boosting on $c_{1b}$ is a resolution of $c_{1b} \gtrsim 0.1$, which motivated our choice
for the cut in Fig. \ref{mtvary}.  The tighter the cut (on either $c_{1b}$ or $m_W$), the stronger the peak at the physical
top mass.   However, one should perform a more detailed simulation to determine which cuts are realistic.

One could also be concerned about how PYTHIA handles the top and W decays, and the effect of PYTHIA on the angular distributions. Using other tools to decay the particles (Madgraph and BRIDGE), we have checked that the distribution is not much affected by the way we simulate the decays\footnote{We thank Lian Tao Wang for pointing out this issue.}.

\begin{figure}[t!]
\centering
\label{ttWHT}
\end{figure}

\section{Conclusions}\label{conclusions}
In this paper we have reviewed how natural EWSB scenarios lead to multi-top final states, and how these could be characterized at the LHC.  We have developed a "topness" measurement procedure, which is able to overcome the combinatoric background,
as well as the SM background.

New physics which provides a new source for multi-top final states, will result in events with many jets and at least one lepton.
If the new physics indeed contains tops it should exhibit an efficiency in the few percent range for reconstructing two
tops in the hadronic channel.  This is our first indiction of topness.
This top reconstruction, including ordering, and requiring more than eight jets, significantly
reduces the main SM background ( \ttb+jets), reducing it by $\sim 10^{-5}$.  Additional cuts on the geometric
variable, $c_{1b}$ or on the mass of the W, can reduce the \ttb+jets by an additional factor of ten.  This allows
for sensitivity to the topness of new physics as long as four top production is in the 1-10 pb range.

A hard cut on $c_{1b}$ or on the mass of the W also allows for an additional litmus test for topness, namely a peak
of the number of events containing two hadronically reconstructed tops at the physical mass of the top, when varying
the mass of the top parameter of the reconstruction.


  

Summarizing, we propose to describe  top-rich new physics by requiring a high efficiency to our cuts and sensitivity to the value of $m_t$. This approach is rather model independent as long as there is a sizable branching ratio to four tops, and it is specially useful in non-boosted tops, i.e. in early LHC physics. Moreover, no b-tagging is required.

\acknowledgments
We are indebted to Kevin Black, Tulika Bose, Adam Martin, Tim Tait, Natalia del Toro and Lian Tao Wang  for many useful discussions.  %
The work of VS and TG is supported partly by NSERC funding. EK is supported by DOE grant DE-FG02-01ER-40676,
NSF CAREER grant PHY-0645456, and also by an Alfred P. Sloan Fellowship.
  


\end{document}